\documentclass[preprint,aps,prb]{revtex4-1}
\usepackage{epsfig}
\usepackage{color}
\begin{document}

\title{Transition of Impurity Nanorod Orientation in Epitaxial 
YBa$_2$Cu$_3$O$_{7-\delta}$ Films on Vicinal Substrates}

\author{Jack J. Shi and Judy Z. Wu}
\affiliation{Department of Physics \& Astronomy, 
The University of Kansas, Lawrence, KS 66045}
\date{December 20, 2011}

\begin{abstract} 
A theoretical study of the structural transition of impurity nanorod 
array in epitaxial YBa$_2$Cu$_3$O$_{7-\delta}$ films on vicinal SrTiO$_3$ 
substrates is presented. Two possible types of film/substrate interface 
were considered with one assuming a complete coherence while the other, 
defective as manifested in presence of anti-phase grain boundaries. Only 
in the former case, the increase of the vicinal angle of the substrate 
leads to a substantial change of the strain field in the film, resulting 
in a transition of the nanorod orientation from the normal to in-plane 
direction of the film. The calculation of the threshold vicinal angle for 
the onset of the transition and the lattice deformation of the film due 
to the inclusion of the impurity nanorods is in very good agreement with 
experimental observations.
\end{abstract}

\maketitle

Nanocomposites, in particular those with the nanostructured second phase 
doped into a primary matrix, have received much attention recently since 
functional performances of the material may be directly enhanced through 
microstructure engineering. Examples of such nanocomposites include 
ferroelectric and multiferroic oxides, magnetic oxides, and superconducting 
cuprates \cite{Haugan,Driscoll,Goyal,Emergo1}. For high-temperature 
superconducting YBa$_2$Cu$_3$O$_{7-\delta}$ (YBCO) films, many experimental 
efforts have been made to generate impurity columnar nanostructures 
in the film to maximize the superconducting critical current ($J_c$) 
in magnetic fields. The doping of insulating impurities such as BaZrO$_3$ 
(BZO), BaSnO$_3$ (BSO), YBa$_2$NbO$_6$, and rare-earth tantalates 
have been identified as promising possibilities for that purpose 
\cite{Driscoll,Goyal,Varanasi,Feldmann,Wee,Harrington}. While extended 
columnar defects provide an increased overall pinning force in the 
superconducting films as they effectively pin magnetic vortices along their 
length, a splay alignment of the columnar defects has been found to reduce 
the possibility of vortex hopping and result in stronger pinning in high 
magnetic fields \cite{Hwa,Civale,Elbaum,Wheeler}. The tuning of the defect 
alignment in YBCO films could therefore provide a possibility to optimize 
the overall behavior of $J_c$ with, especially, reduced $J_c$ anisotropy 
with respect to the direction of magnetic field. In order to achieve the 
controlled alignment of columnar nanostructures such as BZO nanorods in YBCO 
films, we recently employed vicinal SrTiO$_3$ (STO) substrates to introduce 
controllable structural parameters for the formation of nanostructures 
\cite{Baca2,Emergo2010}. It was observed that the orientation of BZO 
nanorod arrays in YBCO films depends on the vicinal angle of the vicinal 
substrate. Increasing the vicinal angle can change the nanorod alignment 
from the normal direction of the film, to splayed around the normal at 
5$^{\circ}$, and to the $ab$ plane of the film at larger angles 
\cite{Baca2,Emergo2010}. In the splayed case, much enhanced overall $J_c$ 
was obtained in magnetic fields up to 5 T with almost any field orientation. 
Theoretical understanding of the role of the vicinal substrate on the 
nanorod orientation is thus important but not currently available. This 
paper presents the first theoretical study on the impurity nanorod alignment 
in epitaxial YBCO films on vicinal substrates. The propose model for the 
formation of nanostructures is rather generic and could be applied to a 
variety of epitaxial nanocomposite films to guide growth of nanostructures 
in a controllable fashion.     

In this study, the effect of vicinal substrates on the formation of 
impurity nanorods in epitaxial YBCO films is modeled in terms of the 
effect of lattice-mismatched substrates. The configurations of the 
nanorods in epitaxial YBCO films on lattice-mismatched substrates 
have been studied with a micromechanical model \cite{shi}, in which 
the formation of the nanorods was assumed to be the consequence of 
the relaxation to the energetically-preferred elastic equilibrium of 
coherently strained lattices due to lattice mismatches among film, 
dopant, and substrate. The possibility to fabricate impurity nanorod 
arrays with the vertical or horizontal alignment in the $c$-oriented 
YBCO films depends on the elastic energy of the strained lattices 
with respect to other possible nanorod configurations. The elastic 
energy of strained lattices is defined as  
\begin{equation}
\label{E_el}
E_{el}=\int\limits_{film}E_1\,dV+\int\limits_{dopant}E_2\,dV
      +\int\limits_{substrate}E_3\,dV
\end{equation}
where $E_i$ is the elastic energy density for film ($i=1$), dopant 
($i=2$), and substrate ($i=3$), respectively. Considering the 
tetragonal symmetry of twinned YBCO film and the cubic symmetry of 
substrate and dopant, the elastic energy density can be written as 
\begin{equation}
\label{E_YBCO1}
E_i=\frac{1}{2}\lambda_{i1} u_{11}^2+\frac{1}{2}\lambda_{i2} u_{33}^2  
    +\lambda_{i3} u_{11}u_{33}+\lambda_{i4} u_{13}^2 
\end{equation} 
where $u_{jk}$ is the strain tensor and $\lambda_{i1}=c^{(i)}_{11}$, 
$\lambda_{i2}=c^{(i)}_{33}$, $\lambda_{i3}=c^{(i)}_{13}$, and 
$\lambda_{i4}=c^{(i)}_{55}$ are the elastic constants of the material 
labeled with $i$. The elastic constants used in this study can be 
found in Refs. \cite{Kuzel,Dieguez,Bouhemadou,Poindexter}. Note that 
the interaction energies at film/substrate and film/dopant interfaces 
are included implicitly in $E_{i}$ as $u_{jk}$ is the solution of 
equilibrium equations with the boundary conditions that are the result 
of the interface interactions. The equilibrium equations can be written 
as 
\begin{equation}
\label{equilibrium}
\sum_{k}\frac{\partial }{\partial x_k}
\left(\frac{\partial E_i}{\partial u_{jk}} \right)=0  \,,
\end{equation}
where $x_i$ is the coordinate along the [100] ($i=1$), [010] ($i=2$), 
and [001] ($i=3$) direction, respectively, and 
$\sigma_{jk}=\partial E_i/\partial u_{jk}$ is the stress tensor. At an 
interface between two coherently bonded lattices, the boundary condition 
of Eq. (\ref{equilibrium}) prescribes continuity of the force on the 
interface and allows for a discontinuity of the strain across the 
interface, $i.e.$ \cite{shi}
\begin{equation}
\label{boundary}
\left\{\begin{array}{l}
{\displaystyle \sum_k n_k \left[\sigma_{jk}(1)-\sigma_{jk}(2)\right] 
   = 0 } \\ 
{\displaystyle \sum_k s_k\left[\sigma_{jk}(1)+\sigma_{jk}(2)\right]  
   = 0 } \\
{\displaystyle \sum_k s_k\left[u_{kk}(1)-(1+f_k)\,u_{kk}(2)-f_k\right]
   = 0 }
\end{array}\right.
\end{equation}
where $\vec{n}$ and $\vec{s}$ are the unit vectors normal and 
tangential to the interface, $(u_{jk}(1),\sigma_{jk}(1))$ and 
$(u_{jk}(2),\sigma_{jk}(2))$ are the strain and stress at the interface 
in lattice $1$ and $2$, respectively, and $f_k$ is the lattice 
mismatches at the interface along the $s_k$ direction. At the top 
surface of a film the boundary condition is simply $n_k\sigma_{jk}=0$ 
and deep inside the substrate the strain vanishes. Consider the case 
that the $c$-oriented YBCO film mismatches with the substrate lattice 
along the [100] direction. To determine the energetically-preferred 
orientation of impurity nanorods in the film, the equilibrium strain 
was solved from Eq. (\ref{equilibrium}) for the nanorods aligned in the 
[001] or [100] direction of the film and the difference of the elastic 
energies with the two alignments were obtained as \cite{shi}
\begin{equation}
\label{dE1}
E_{el}[100]-E_{el}[001]
 =E_0 \left[\left(1-\rho+\frac{w_1}{w_2}\rho\right)G
-\frac{w_1\alpha(1-\alpha)}{w_2}\rho \,\right] \,, 
\end{equation}
where $E_{el}[100]$ and $E_{el}[001]$ are the elastic energies of the 
film with the nanorods aligned in the $[100]$ and $[001]$ directions, 
respectively, $E_0$ is a positive function, $\rho$ is the volume 
density of nanorods, 
\begin{equation}
w_i=\lambda_{i1}-\lambda_{i3}^2/\lambda_{i2}
\end{equation}
is of the elastic constants of the film ($i=1$), dopant ($i=2$), and 
substrate ($i=3$), respectively, and $\alpha=\lambda_{12}/\lambda_{11}$. 
In Eq. (\ref{dE1}), 
\begin{equation}
\label{phase}
G=\left[\frac{(\alpha w_1+w_2)(f_1+\delta f_1)}
             {(w_1+w_2)(f_3+\delta f_3)}\right]^2-\alpha\,,
\end{equation}
where $f_1+\delta f_1$ and $f_3+\delta f_3$ are the lattice mismatches 
between the dopant and the strained film lattice due to the mismatched 
substrate. These lattice mismatches were calculated in Ref. \cite{shi} 
as
\begin{eqnarray}
f_1+\delta f_1 &=& \frac{a_2}{a_1(1+\Gamma)}-1 \,,\\
f_3+\delta f_3 &=& \frac{c_2}{c_1(1-\lambda_{13}
                   \Gamma/\lambda_{12})}-1 \,,
\end{eqnarray} 
where $(a_1,c_1)$ and $(a_2,c_2)$ are the lattice constants of the 
film and dopent along [100] and [001] direction, respectively,
\begin{equation}
\label{Gamma}
\Gamma =\frac{w_3f_s}{w_3+w_1(1+f_s)}\,,
\end{equation} 
$f_s=a_3/a_1-1$ is the lattice match between the film and substrate 
along the [100] direction, and $a_3$ is the natural lattice constant
of the substrate. For $\rho<<1$, $G$ as a function of $w_2$, $f_1$, 
$f_3$, and $f_s$ can be conveniently used as a state function for 
the nanorod orientation. When $G<0$, $E_{el}[100]-E_{el}[001]<0$ and 
it is not possible to have nanorods aligned in the [001] direction 
and vice versa. Hence, $G=0$ yields a phase boundary that separates 
the regions in the parameter space of $(w_2,f_1,f_3,f_s)$ where the 
vertical or horizontal alignment of nanorods is not possible in the 
$c$-oriented film on a lattice-mismatched substrate. For a given type
of impurity doping in the YBCO films, therefore, the tuning of the 
substrate lattice constant, if possible, could result in a transition 
of the nanorod orientation \cite{shi}. 

For YBCO epitaxy films on the vicinal substrate, two possible YBCO 
crystalline configurations near the film/substrate interface have been 
observed experimentally. In the first configuration (see Fig. 1a), the 
YBCO $ab$ planes follow the substrate (001) planes on terraces (miscut 
steps) on the vicinal substrate surface and antiphase boundaries are 
typically formed in YBCO along the $c$ axis at (or near) the step 
edges of the terraces \cite{Wen,Haage,Zegen}. In this case, the YBCO 
$ab$ planes can become detwinned with the shorter $a$ axis of the 
orthorhombic YBCO layer preferably oriented along the miscut direction 
\cite{Brotz,Dekkers}. The percentage of the untwinning was observed 
to increase linearly with the vicinal angle \cite{Brotz} and almost 
complete (70\% to 90\%) detwinning occurs at $10^{\circ}$ vicinal angle 
in one study \cite{Brotz} and less than $1^{\circ}$ vicinal angle in 
another study \cite{Dekkers}. In the second observed crystalline 
configuration (see Fig. 1b), the YBCO $c$ axis aligns with the normal 
of the substrate optical surface and the YBCO $ab$ planes follow the 
substrate surface \cite{Dekkers,Chakalova,Maurice1,Maurice2}. In 
this case, very few antiphase boundaries were observed in the film 
\cite{Dekkers,Chakalova} and a smooth and homogeneous surface 
morphology was observed by the AFM imaging \cite{Chakalova}. Such a
configuration has also been observed in YBCO films on vicinal MgO 
substrates \cite{Streiffer}. For the doped YBCO films with impurity 
nanorods on vicinal substrates, it is not clear which crystalline 
configuration dominates near the film/substrate interface, although 
the lack of the antiphase boundary in the samples \cite{Javier} hints 
the possibility of the second configuration. To study the nanorod 
configurations, both the crystalline possibilities of the YBCO 
over-layers near the substrate surface were considered in this work. 
In the first possible crystalline configuration, the formation of the 
antiphase boundaries in the YBCO lattice could change the elastic 
constants as well as the strain field of the YBCO lattice. As very few 
antiphase boundaries have been observed in the YBCO film with impurity 
nanorod inclusions, the effects of the antiphase boundary were not 
considered here. To model the effect of the detwinning of the YBCO 
$ab$ planes, the lattice constant of the YBCO $a$ axis is assumed to 
decrease linearly from the average lattice constant in the $ab$ plane 
($\bar{a}=3.855$ \AA) to its natural value ($a_{01}=3.823$ \AA) as 
the vicinal angle $\phi$ increases, $i.e.$ 
\begin{equation}
\label{a1}
a_1=a_{01}+(\bar{a}-a_{01})zH(z)
\end{equation}
where $z=1-\phi/\phi_u\,$, $\phi_u$ is the vicinal angle at which the 
$ab$ plane is completely detwinned, and $H(z)$ is the Heaviside step 
function as $H(z)=0$ for $z<0$ and $H(z)=1$ for $z\geq 0$. In the case 
that the YBCO $c$ axis is perpendicular to the optical surface of the 
vicinal substrate, the effective lattice constant of the substrate is 
\begin{equation}
\label{a3}
a_3=\frac{a_{03}}{1-\chi(1-\cos\phi)}
\end{equation} 
where $a_{03}$ is the natural lattice constant of the substrate and 
$\chi$ is a parameter for easily switching the direction of the film 
lattice in the calculation. The film $ab$ planes are parallel to 
the substrate lattice if $\chi=0$ and with the substrate optical 
surface if $\chi=1$. Including both effects of the detwinning and 
tilting of the film $ab$ planes, the lattice mismatch between the 
film and substrate along the miscut direction ([100] direction) is 
\begin{equation}
\label{mismatch0}
f_s=\frac{a_{03}}
    {[1-\chi(1-\cos\phi)][a_{01}+(\bar{a}-a_{01})zH(z)]}-1
\end{equation} 
In the direction perpendicular to the miscut direction, the mismatch 
between YBCO and STO substrate at the film/substrate interface is 
negligible. Substituting $f_s$ in Eq. (\ref{mismatch0}) into 
Eqs. (\ref{phase})--(\ref{Gamma}) yields $G$ as a function of the 
vicinal angle $\phi$ for the orientation of impurity nanorods in the 
$c$-oriented YBCO film on the vicinal substrate. Figure 2 plots 
$G=G(\phi)$ for two possible YBCO crystalline configurations on the 
vicinal substrate. Since $G(\phi)>0$ when $\chi=0$, the detwinning of 
the YBCO $ab$ planes is not the root cause of the observed transition 
of the nanorod orientation. In fact, the curves with different 
$\phi_u$ overlaps in Fig. 2 and, therefore, the change of the strain 
field in the film due to the detwinning has no significant effect on 
the nanorod orientation. When the YBCO $ab$ planes are parallel to 
the substrate optical surface ($\chi=1$), $G(\phi)$ changes from 
positive to negative at a threshold vicinal angle as shown in Fig. 2. 
This is when the lattice mismatches between YBCO and dopant are 
substantially altered by the strain in YBCO due to the lattice 
mismatch with the substrate and, consequently, the 
energetically-preferred nanorod alignment switches from the [001] to 
[100] direction. From Fig. 2, this threshold angle for the onset of 
the transition of the nanorod orientation is found to be 
$\phi_c\simeq 5.5^{\circ}$ for both the BZO and BSO nanorods. 
Experimentally \cite{Baca2}, it was observed that the orientation of 
the nanorods is splayed at $\phi=5^{\circ}$ and changes into the [100] 
direction when $\phi\geq 10^{\circ}$, which is in general consistent 
with the model prediction. Since $\phi=5^{\circ}$ is very close to 
$\phi_c$, the difference between the elastic energies of the 
horizontal and vertical alignments is very small and the nanorods 
could become splayed. The model calculation has therefore suggested 
that near the film/substrate interface the film $ab$ planes are 
parallel with the substrate optical surface and the strain due to the 
lattice mismatch between the film and the tilted substrate lattice 
is the cause of the transition of the nanorod orientation.

The deformation of the film lattice due to the inclusion of 
the nanorods can be calculated by averaging the principal components 
of the equilibrium strain over the film \cite{shi}, which can be 
compared with experimental measurement. Because of the different 
nanorod orientation, the deformation of the YBCO lattice is different 
in the regions of $\phi<\phi_c$ and $\phi>\phi_c$. When $\phi<\phi_c$, 
the deformation of YBCO calculated from the equilibrium strain is 
\cite{shi}
\begin{equation}
\label{Da3}
\begin{array}{l}
{\displaystyle
\frac{\delta a_1}{a_1}=-\frac{\lambda_{13}w_2(f_3+\delta f_3)}
      {2\,[\lambda_{11}w_2+(1+f_3+\delta f_3)\lambda_{12}w_1]} 
      +\frac{1}{2}\Gamma }     
   \vspace{0.1in} \\
{\displaystyle
\frac{\delta c_1}{c_1}=\;\;\;\frac{\lambda_{11}w_2(f_3+\delta f_3)}
      {2\,[\lambda_{11}w_2+(1+f_3+\delta f_3)\lambda_{12}w_1]}
      -\frac{\lambda_{13}}{2\lambda_{12}}\Gamma  }
\end{array}
\end{equation}      
For BZO or BSO nanorods in YBCO films on STO substrates, 
$f_3+\delta f_3>0$. Hence $\delta a_1<0$ and $\delta c_1>0$, which 
represents a compression and expansion of the YBCO lattice along the 
$a$ and $c$ axis, respectively. When $\phi>\phi_c$, the deformation 
of the YBCO lattice is calculated as \cite{shi}
\begin{equation}
\label{Da4}
\begin{array}{l}
{\displaystyle \frac{\delta a_1}{a_1} 
 =\frac{1}{2}\left[ \frac{w_2(f_1+\delta f_1)}
                         {w_2+(1+f_1+\delta f_3)w_1}+\Gamma \right]} 
                         \vspace{0.1in} \\
{\displaystyle \frac{\delta c_1}{c_1}
 =-\left(\frac{\lambda_{13}}{\lambda_{12}}\right)
   \left(\frac{\delta a_1}{a_1}\right) }                         
\end{array}
\end{equation}
In this case, the YBCO lattice is expanded along the $a$ axis and 
compressed along the $c$ axis. Note that the ratio of the film lattice 
deformations along the [001] and [100] direction is independent of the 
properties of the dopant and substrate. Figure 3 plots $\delta a_1/a_1$ 
and $\delta c_1/c_1$ as functions of $\phi$ calculated from 
Eqs. (\ref{Da3}) and (\ref{Da4}) with $f_s$ from Eq. (\ref{mismatch0}) 
for the case of BZO nanorods in the c-oriented YBCO films on vicinal 
STO substrates. The currently available experimental data from the 
X-ray measurements of the $c$-axis lattice constant \cite{Emergo} was 
also included in the figure. Overall, the model calculation agrees well 
with the experimental data. The discrepancy between the experiment and 
the model when $\phi>15^{\circ}$ is likely due to the incoherent growth 
of YBCO on the vicinal substrate. At a relatively large vicinal angle, 
the lattice mismatch between YBCO and substrate is too big ($\sim 6.5\%$ 
at $\phi=20^{\circ}$) to allow a coherent film growth. If incoherent 
lattice growth prevails, the strain in the lattice will be released. 
This is similar to the case of porous YBCO films on vicinal substrates,
in which a large number of pores appear at approximately 10-15 nm away 
from the substrate/film interface \cite{Wu1,Emergo3}. To further confirm 
the model prediction on the lattice deformation due to the inclusion of 
impurity nanorods, experimental studies of the lattice deformation in 
the film $ab$ planes are needed.  

In conclusion, a micromechanical model based on the theory of elasticity 
has been developed to study the configurations of the self-organized 
impurity nanorod array in $c$-oriented YBCO films on vicinal STO 
substrates. By treating lattice mismatch locally at the interfaces, the 
strain field due to multiple mismatched lattices of film, dopants, and 
substrate can be considered simultaneously. Including the effect of 
multiple lattice mismatches is important to the understanding of impurity 
nanostructures in nanocomposite films when multiple lattices are involved.  
In the case of BZO doped YBCO films on vicinal substrates, two possible 
types of film/substrate interface were studied: perfectly coherent and 
quasi-coherent. The former assumes negligible defects or dislocations on 
the interface, which results in a significantly strained film lattice. 
The increase of the lattice strain in the film with the vicinal angle of 
the substrate leads to a transition of the nanorod orientation from the 
normal to the parallel direction of the film. The calculated threshold 
vicinal angle for the onset of the transition for BZO nanorods and the 
predicated deformation of the YBCO lattice due to the inclusion of the 
nanorods are in very good agreement with the experimental observations. 
This agreement is significant considering no fitting parameter was 
employed in the model and the approach may apply to many other 
nanocomposites in design and synthesis of novel nanostructures. 

This work is supported by NSF and ARO under contract no. 
NSF-DMR-0803149, NSF-DMR-1105986, NSF EPSCoR-0903806, and 
ARO-W911NF-0910295.

\newpage

\begin{center} 
\includegraphics[width=70mm,angle=0]{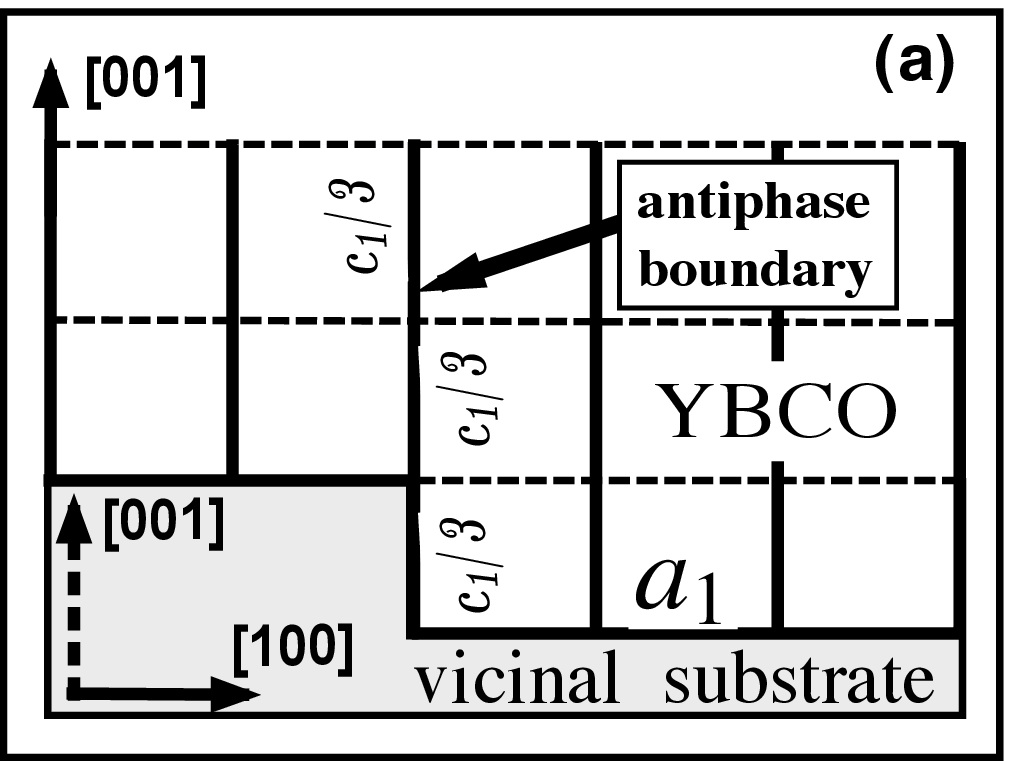}
\hspace{0.2in}
\includegraphics[width=70mm,angle=0]{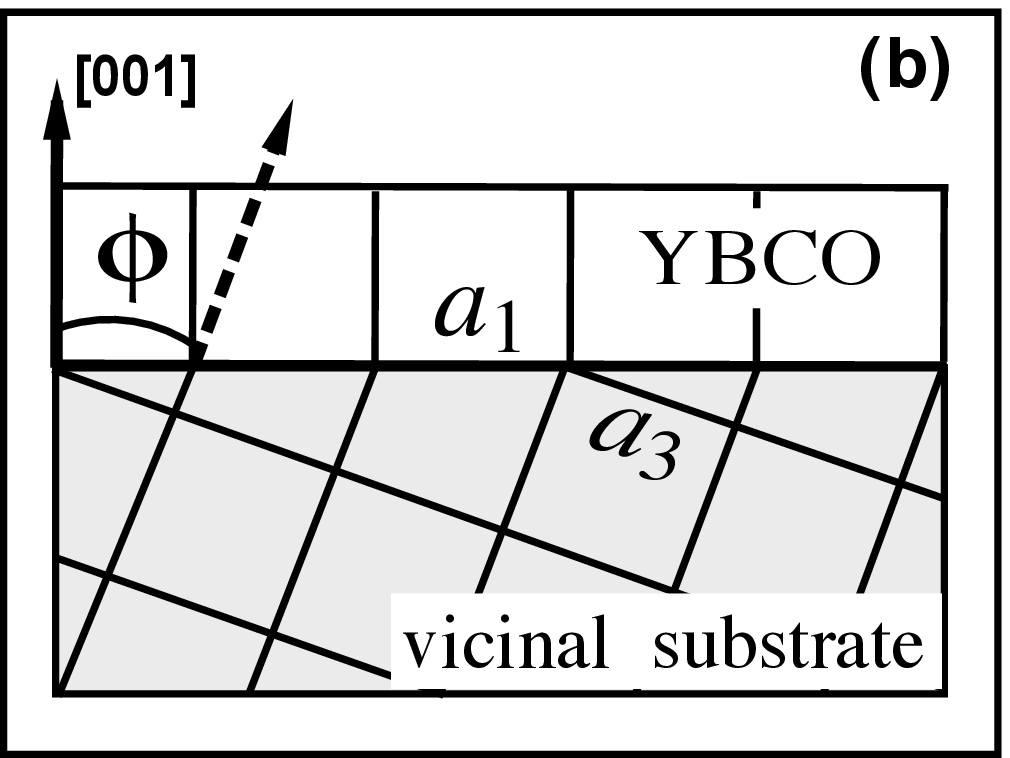}

\vspace{0.2in}
\begin{minipage}{6.0in}
Figure 1. Illustration of possible YBCO crystalline configurations
near the film/substrate interface on a vicinal STO substrate, where
the YBCO $ab$ planes parallel to (a) the substrate (001) planes and 
(b) the substrate optical surface. The dashed and vertical solid 
arrows indicate the [001] direction of the substrate and film, 
respectively. 

\end{minipage}
\end{center}

\vspace{0.2in}
\begin{center} 
\includegraphics[width=80mm,angle=-90]{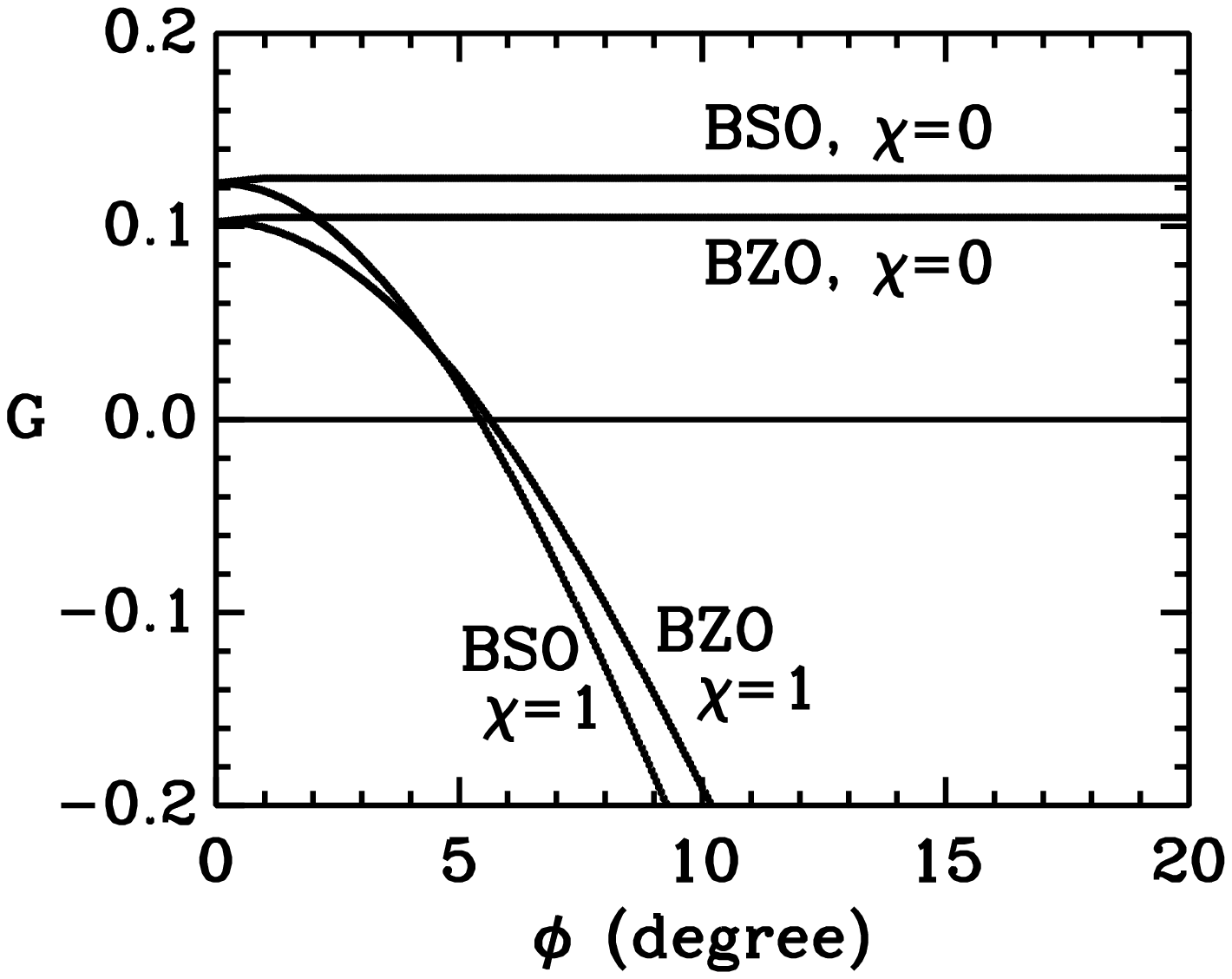}

\vspace{0.2in}
\begin{minipage}{6.0in}
Figure 2. 
$G$ $v.s.$ vicinal angle for BZO or BSO nanorods in YBCO films on vicinal 
STO substrates. Curves with only difference in $\phi_u$ are overlapped.

\end{minipage}
\end{center}

\vspace{0.2in}
\begin{center} 
\includegraphics[width=80mm,angle=-90]{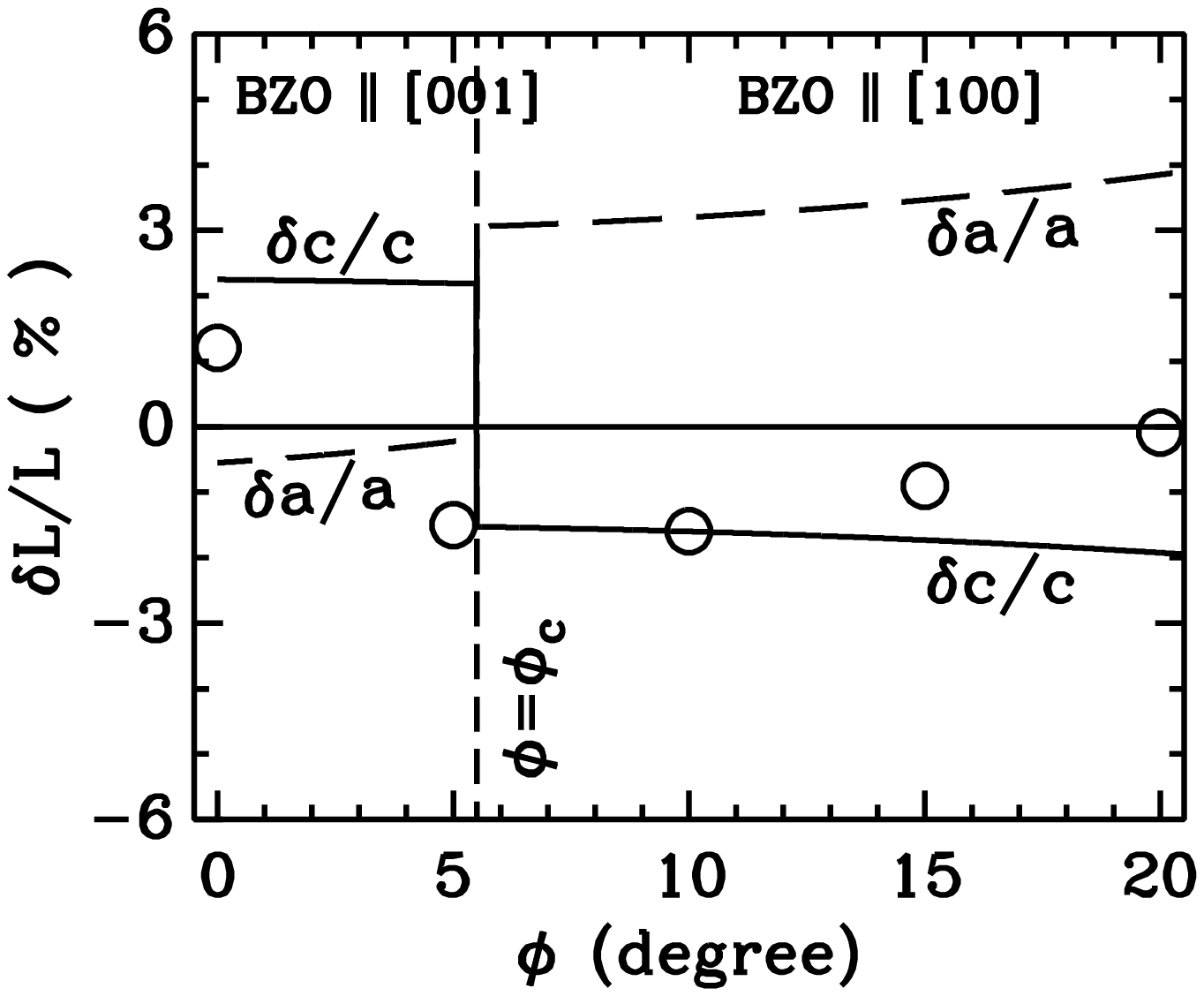}

\vspace{0.2in}
\begin{minipage}{6.0in}
Figure 3. 
Calculated percentage change of YBCO lattice constants ($\delta L/L$)
as a function of vicinal angle ($\phi$) for BZO nanorods in YBCO films
on vicinal STO substrates. The solid and dashed curve are of the $c$ 
and $a$ axis, respectively. Open circles are of the X-ray measurement 
of the $c$-axis. The vertical dashed line marks the threshold vicinal 
angle for the onset of the transition of the nanorod orientation. 
\end{minipage}
\end{center}


\begin{thebibliography}{99}
\bibitem{Haugan} 
T.J. Haugan, $et.$ $al.$, Nature {\bf 430}, 867 (2004).

\bibitem{Driscoll} 
J.L. MacManus-Driscoll, $et.$ $al.$, Nature Mater. {\bf 3}, 439 (2004).

\bibitem{Goyal}    
A. Goyal $et.$ $al.$, Supercond. Sci. Technol. {\bf 18}, 1533 (2005). 

\bibitem{Emergo1}  
R. Emergo, J.Z. Wu, T. Haugan, and P. Barnes, 
Appl. Phys. Lett. {\bf 87}, 232503 (2005).

\bibitem{Varanasi} 
C.V. Varanasi, $et.$ $al.$, J. Appl. Phys. {\bf 102}, 063909 (2007).

\bibitem{Feldmann} 
D.M. Feldmann, $et.$ $al.$, Supercond. Sci. Technol. {\bf 23}, 
095004 (2010).

\bibitem{Wee} 
S.H. Wee, $et.$ $al.$, Appl. Phys. Express {\bf 3}, 023101 (2010).

\bibitem{Harrington} 
S.A. Harrington, $et.$ $al.$, Supercond. Sci. Technol. {\bf 22}, 
022001 (2009).

\bibitem{Hwa} 
T. Hwa, P. Le Doussal, D.R. Nelson, and V.M. Vinokur, 
Phys. Rev. Lett. {\bf 71}, 3545 (1993).

\bibitem{Civale} 
L. Civale, $et.$ $al.$, Phys. Rev. B{\bf 50}, 4102 (1994).

\bibitem{Elbaum} 
L. Krusin-Elbaum, $et.$ $al.$, Phys. Rev. Lett. {\bf 76}, 2563 (1996).

\bibitem{Wheeler} 
R. Wheeler, $et.$ $al.$, Appl. Phys. Lett. {\bf 63}, 1573 (1993).

\bibitem{Baca2}
F.J. Baca $et.$ $al.$, Appl. Phys. Lett. {\bf 94}, 102512 (2009)

\bibitem{Emergo2010}
R. Emergo, J. Baca, J.Z. Wu, T. Haugan and P. Barnes, 
Supercond. Sci. Technol. {\bf 23}, 115010 (2010).

\bibitem{shi}
J. Shi and J.Z. Wu, ``Micromechanical Model for Self-Organized Impurity 
Nanorod Arrays in Epitaxial YBa$_2$Cu$_3$O$_{7-\delta}$ Films'',
preprint (2011).

\bibitem{Kuzel}     
P. Kuzel $et.$ $al.$, J. Phys.: Condens. Matter {\bf 13}, 167 (2001).
\bibitem{Dieguez}   
O. Di\'{e}guez, K.M. Rabe, and D. Vanderbilt, Phys. Rev. B{\bf 72}, 
144101 (2005).
\bibitem{Bouhemadou}  
A. Bouhemadou and K. Haddadi, Solid State Sciences {\bf 12}, 630 (2010).
\bibitem{Poindexter}  
E. Poindexter and A.A. Gardini, Phys. Rev. {\bf 110}, 1069 (1958).

\bibitem{Wen}
J.G. Wen, C. Traeholt, and H.W. Zandbergen, Physica C{\bf 205}, 354 (1993).
\bibitem{Haage}
T. Haage, J. Zegenhagen, H.-U. Habermeier, and M. Cardona, 
Phys. Rev. Lett. {\bf 80}, 4225 (1998).
\bibitem{Zegen}
J. Zegenhagen, T. Haage, and Q.D. Jiang, Appl. Phys. A{\bf 67}, 711 (1998).
\bibitem{Brotz}
J. Br\"{o}tz, H. Fuess, T. Haage, and J. Zegenhagen, 
Phys. Rev. B{\bf 57}, 3679 (1998).
\bibitem{Dekkers}
J.M. Dekkers, $et.$ $al.$, Appl. Phys. Let. {\bf 83}, 5199 (2003).
\bibitem{Chakalova} 
R.I. Chakalova, $et.$ $al.$, Phys. Rev. B{\bf 70}, 214504 (2004).  
\bibitem{Maurice1}
J.L. Maurice, O. Durand, M. Drouet, and J.-P. Contour,
Thin Solid Films {\bf 319}, 211 (1998).
\bibitem{Maurice2}
J.L. Maurice, O. Durand, K. Bouzehouane, and J.-P. Contour,
Physica C {\bf 351}, 5 (2001).
\bibitem{Streiffer}
S.K. Streiffer, B.M. Lairson, and J.C. Bravman, Appl. Phys. Lett. 
{\bf 57}, 2501 (1990).
\bibitem{Javier}
F.J. Baca, 
``In-Situ Control of BaZrO$_3$ and BaSnO$_3$ Nanorod Alignment and 
Microstructure in YBa$_2$Cu$_3$O$_{7-x}$ Thin Films by Strain 
Modulated Growth'', Ph.D. Dissertation, University of Kansas, (2009).
\bibitem{Emergo}
R. Emergo, J.Z. Wu, T.J. Haugan, and P.N. Barnes, 
Supercond. Sci. Technol. {\bf 21}, 085008 (2008).
\bibitem{Wu1}
J.Z. Wu, $et.$ $al.$, Appl. Phys. Lett. {\bf 93}, 062506 (2008).
\bibitem{Emergo3}
R. Emergo, J.Z. Wu, D.K. Christen and T. Aytug, 
Appl. Phys. Lett. {\bf 85}, 70 (2004).
\end{thebibliography}
\end{document}